\documentclass[aps,twocolumn,prb,amsmath,amssymb,floatfix]{revtex4}
\usepackage{graphicx}
\usepackage{amssymb,amsmath,latexsym}
\usepackage{bm,color}

\begin{document}

\def\cblue{\color{blue}}
\def\cred{\color{red}}  

\title{The role of Coulomb correlation in charge density wave of CuTe}
\author{Sooran Kim$^1$}
\author{Bongjae Kim$^{2,3}$}
\author{Kyoo Kim$^3$}
\email[]{kyoo@mpk.or.kr}
\affiliation{
$^1$Department of Physics Education, Kyungpook National University, Daegu 41566, Korea \\ 
$^2$Department of Physics, Kunsan National University, Gunsan 54150, Korea \\ 
$^3$Max Planck POSTECH/Hsinchu Center for Complex Phase Materials,
        Pohang University of Science and Technology, Pohang 37673, Korea \\
}
\date{\today}


\begin{abstract}
A quasi one-dimensional layered material, CuTe undergoes a charge density wave (CDW) transition in Te chains with a modulation vector of $q_{CDW}=(0.4, 0.0, 0.5)$. 
Despite the clear experimental evidence for the CDW, the theoretical understanding especially the role of the electron-electron correlation in the CDW has not been fully explored. Here, using first principles calculations, we demonstrate the correlation effect of Cu is critical to stabilize the 5$\times$1$\times$2 modulation of Te chains. 
We find that the phonon calculation with the strong Coulomb correlation exhibits the imaginary phonon frequency so-called phonon soft mode at $q_{ph0}=(0.4, 0.0, 0.5)$ indicating the structural instability. 
The corresponding lattice distortion of the soft mode agrees well with the experimental modulation.
These results demonstrate that the CDW transition in CuTe originates from the interplay of the Coulomb correlation and electron-phonon interaction.
\end{abstract}


\maketitle

\section{ Introduction }
The novel electronic and magnetic properties of low dimensional materials have drawn interest because of their fundamental physics and possible applications\cite{Stormer99,Zeng08,Kenneth18,Kim18,Gibertini19,Chia18}.
The intrinsic instabilities in low dimensional systems often trigger charge density wave (CDW), Peierls transitions, spin density wave, or even unconventional superconductivity\cite{Zhu15,Zhu17,Gabovich01,Lee06,Cao18}. 
Peierls-type transitions are experimentally observed in quasi one-dimensional (1D) materials such as 
TTF-TCNQ molecular solid {\it trans-}polyacetylene polymers, ${\it MQ}_3$ ({\it M}=Ta, Nb and {\it Q}=S, Se),  K$_{0.3}$MoO$_{3}$
\cite{Denoyer75,Kagoshima75,Jerome04, Sambongi77,Smaalen05,Wang89,Boswell80, Pouget85, Fleming85} 
(see Table I within Ref. \onlinecite{Smaalen05}).

Electronic instability, however, widely understood as the origin of a Peierls transition, has been challenged\cite{Mazin08, Zhu15, Zhu17, Kartoon18}.
As the dimension of interatomic connection increases, the susceptibility 
peak feature becomes weaken, and other mechanisms such as electron-phonon interaction become important in the realization of a Peierls-type structural, or CDW transition. 
Sometimes, the role of the underlying 1D interatomic network can be 
pronounced due to the directional bonding, for instance, of $p$ orbitals in elements 
such as Se, Te, and I, resulting in strong 1D Peierls-type structural transitions in higher dimensions\cite{Decker02, Min06}.
Furthermore, even though the Peierls transition does not require the strong electron-electron correlation as in the Mott transition, there have been reports on the role of strong Coulomb correlation in the Peierls transition, dubbed as a Mott-Peierls transition, especially, in VO$_2$\cite{Jeckelmann98,Sengupta03,Biermann05,Haverkort05,Kim13,Kim14}. 
Therefore elucidating the mechanism of CDW transitions, which can be originated from the Fermi surface nesting, electron-phonon interaction, electron-electron correlation, or even the interplay of them would be interesting and important for fundamental physics in low dimensional systems. 
\\

\begin{figure}[b]
\includegraphics[width=6.5cm]{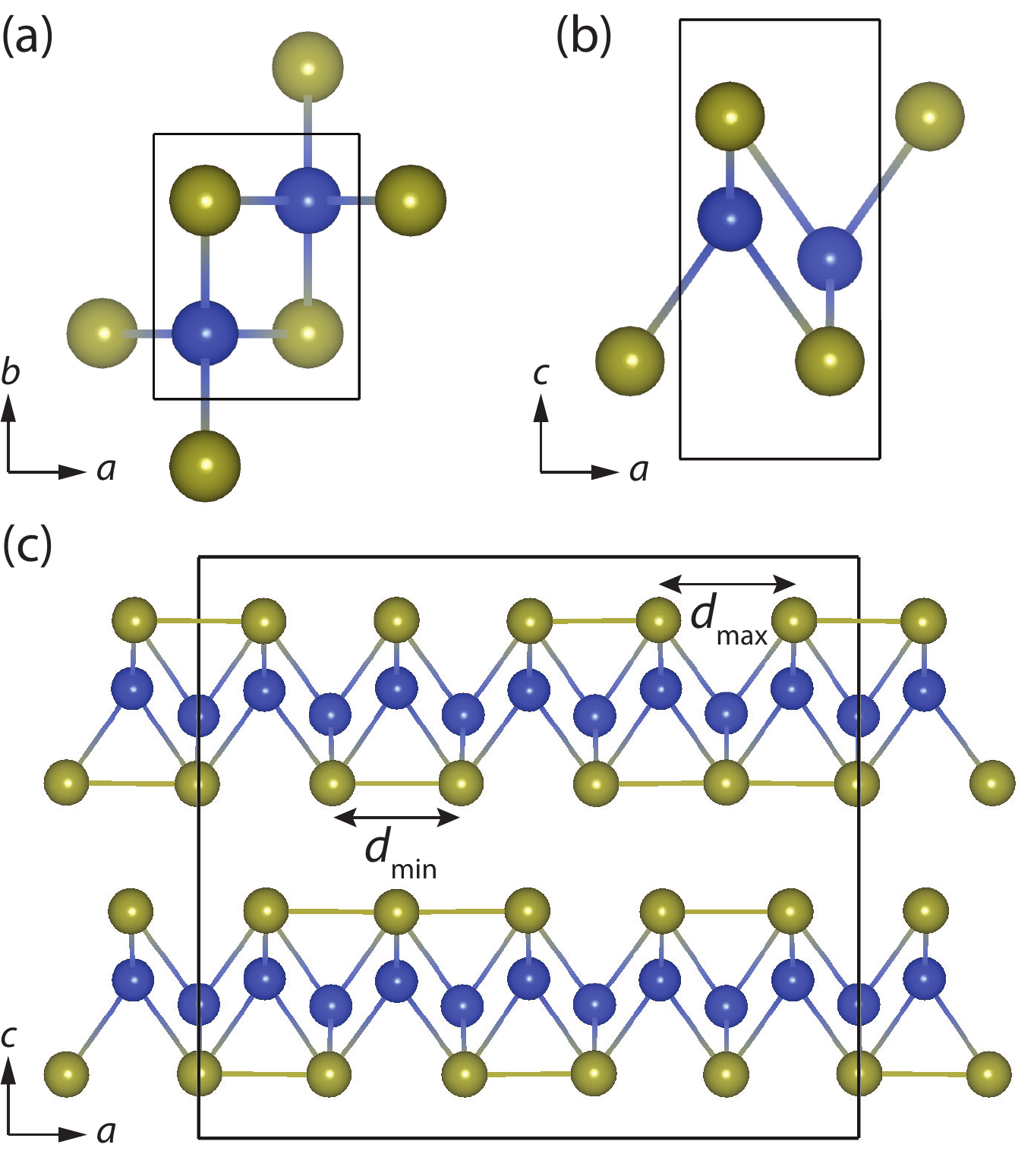}
\caption{Crystal structures of CuTe. Blue and yellow balls represent Cu and Te, respectively.
        (a) high symmetric structure in the non-CDW phase at a high temperature.
        (b) 5$\times$1$\times$2 modulated structure in the CDW phase. The bond between Te atoms illustrates the Te modulation along the $a$ direction reported in REF. \onlinecite{Stolze13,Zhang18}. $d_{min}$ and $d_{max}$ indicate the shortest and longest distances among Te-Te bondings, respectively.
}
\label{str}
\end{figure}
 
The crystal CuTe, called vulcanite, is one of the prototypical quasi 1D systems, which undergoes a CDW transition at $T_{CDW}=$ 335 K.
The early X-ray diffraction study reported that CuTe is crystallized in the strained 
FeTe-like orthorhombic unit cell with the space group {\it Pmmn} (No. 59) which consists of one formula unit of CuTe \cite{Pertlik01}. 
As shown in Fig. \ref{str}(a) and (b), Te atoms form a distorted square planar net 
resulting in the quasi 1D chain structure, and Cu atoms have a planar square network 
with buckling in the non-CDW phase of CuTe.

According to the tight binding calculation of Seong {\it et al.}, 
a dimer formation with $q=(0.5,0.0,0.0)$ is stabilized over the non-dimerized state and opens a band gap, 
which suggests the possibility of the structural transition accompanying a metal-insulator transition\cite{Seong94}.
More recent X-ray diffraction, as well as a high-resolution tunneling electron microscopy 
experiment, observed a structural modulation of the Te chain with $q_{CDW}=(0.4,0.0,0.5)$ as shown in Fig. \ref{str}(c)\cite{Stolze13}.
The CDW transition in CuTe is also investigated utilizing angle-resolved photoemission spectroscopy (ARPES) and analyzed with first principles calculations\cite{Zhang18}.
The momentum-dependent gap opening of 0.1-0.2 eV for a quasi 1D band is clearly observed below $T_{CDW}$ in the ARPES signals.
They also demonstrated the band structure evolution with temperature from 20 K to 350 K and potassium doping, and eventual disappearance of the CDW gap feature. 
Both Fermi surface nesting and electron-phonon coupling were reported as an origin of the CDW instability from a peak feature in the bare charge susceptibility and a Kohn anomaly in the phonon calculation at $q_{CDW}$.
Their phonon dispersion curve, however, does not show the imaginary phonon frequency at $q_{CDW}$, which is the evidence of the structural instability\cite{Zhang18}. 

To unveil the microscopic mechanism of the CDW transition, in this paper, we present the electronic structure and lattice dynamics of CuTe  by first principles calculations.
Especially, we focus on the Coulomb correlation of Cu ion because of its partial occupied $d$ orbitals. 
We considered various types of van der Waals interaction scheme, exchange-correlation functionals, and electron-electron correlation strength to explore the origin of the modulation.
Among them, we find that the strong Coulomb correlation of Cu $d$ orbitals has an essential role in triggering the CDW transition. The phonon dispersion curve with considering the correlation effect provides the imaginary frequency whose corresponding lattice displacement is exactly consistent with the experimental Te modulation. 

\section{Computational details}
Density functional theory (DFT) calculations for structural relaxations and force calculations were performed by 
the Vienna {\it ab initio} simulation package, VASP \cite{Kresse96}. 
PHONOPY were used for phonon calculations\cite{Togo08}.  
FPLO was employed to analyze the detailed band structure including band unfolding\cite{Ku10}. 

We included the spin-orbit coupling (SOC) and utilized two exchange-correlation functionals: Perdew-Burke-Ernzerhof functional (PBE)\cite{PBE} and PBEsol (revised PBE for solid)\cite{PBEsol}. 
We performed the PBE+$U$ calculations to account for the correlated $d$ orbitals of Cu with the Dudarev implementation \cite{Dudarev98}. 
$U_{eff}=U-J$ in a range of 2 eV to 13 eV is tested to investigate the Coulomb correlation effect on the structural instability.
Three different types of van der Waals interaction schemes are also checked: DFT-D3 method with zero-damping (D3)\cite{Grimme10} and Becke-Jonson damping (D3-BJ)\cite{Grimme11}, and D2 method of Grimme (D2)\cite{Grimme06}. 
The energy cut for the plane waves in the overall calculation is 400 eV.
For the structural relaxations, the {\bf k}-point sampling for the non-CDW and the CDW structure are 20$\times$16$\times$8 and 4$\times$16$\times$4, respectively.

For the phonon calculation, the dynamical matrix is obtained with finite displacements method (frozen phonon method)
using the 10$\times$1$\times$2 supercell, based on the Hellmann-Feynman theorem\cite{Togo08,Parlinski97}. 
Before carrying out the phonon calculations, we performed the atomic relaxation using experimental 
lattice parameters\cite{Stolze13}. 
The {\bf k}-point sampling of 3$\times$16$\times$4 is used for the 10$\times$1$\times$2 supercell. 

To obtain a reasonable range of Coulomb correlation parameters of Cu atoms, we have employed the linear response method \cite{Cococcioni05} implemented in QUANTUM ESPRESSO\cite{QE09}. 
The dense (48$\times$36$\times$24) {\bf k} mesh is used for the high symmetric primitive unit cell.
The energy cut for wavefunctions and the kinetic energy cut for charge density and potential are 45 Ry and 250 Ry, respectively.

\section{Results and discussions}
\begin{figure}[b]
\includegraphics[width=7cm]{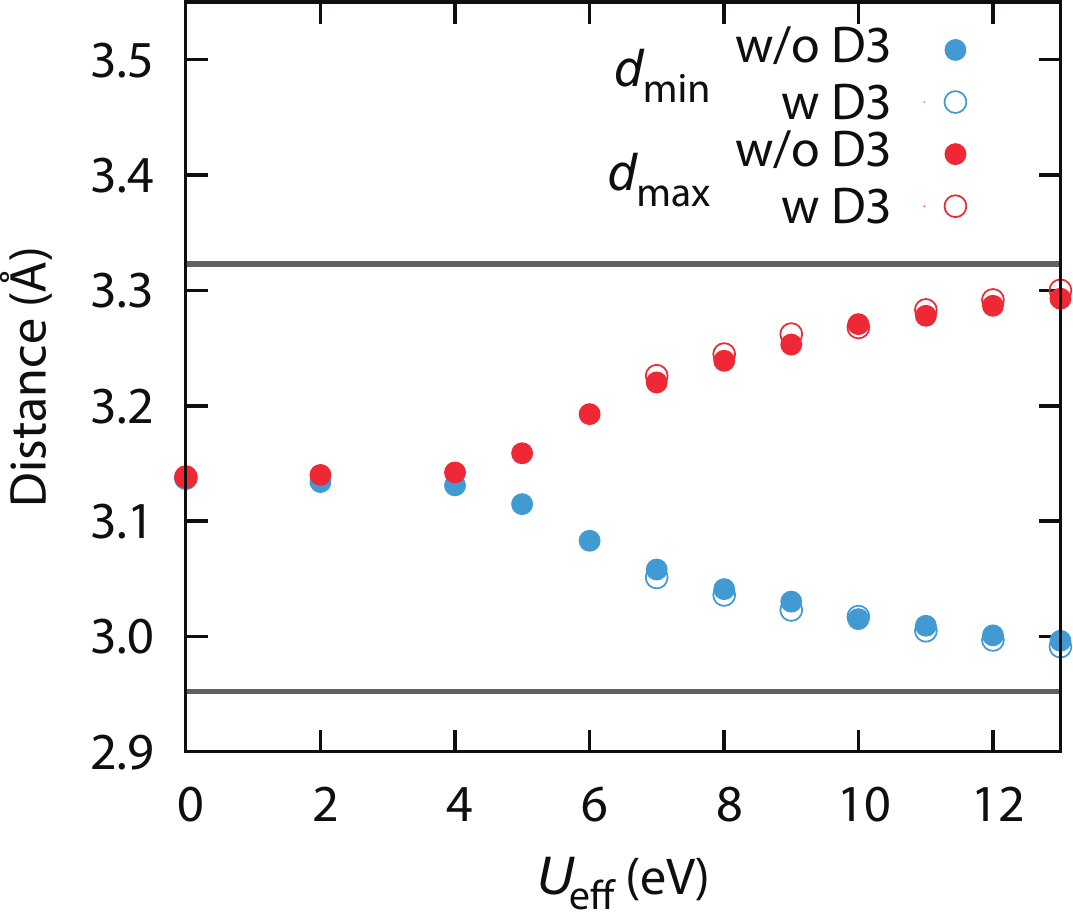}
\caption{Distances of the Te-Te bonding as a function of $U_{eff}$. The solid and dotted lines represent the distances after the relaxation without and with the van der waals interaction with the D3 method, respectively. Gray lines indicate the experimental values.
}
\label{dist}
\end{figure}

To investigate the instability of the Te chains, we relaxed the internal parameters of the 5$\times$1$\times$2 supercell, starting from the experimental one\cite{Stolze13} and obtained $d_{min}$ and $d_{max}$ (See Fig. \ref{str}) in Te chains in various simulation conditions employing diverse types of van der Waals interactions and functional, varying Coulomb correlation parameters ($U_{eff}$), and for hole doping case.
The modulated CDW structure is relaxed back to the high-symmetric non-CDW structure losing the formation of the Te modulation except when the Coulomb correlation for Cu $d$ orbitals is considered. This result is consistent with the stable phonon dispersion curve of Zhang {\it et al.}\cite{Zhang18}, where strong Coulomb correlation is not included.

Figure \ref{dist} shows the calculated $d_{min}$ and $d_{max}$ depending on the $U_{eff}$. 
As $U_{eff}$ increases, clear bifurcation of $d_{min}$ and $d_{max}$ is observed, and their values are progressively reaching the experimental values regardless of the van der Waals correction.
When the $U_{eff}$ is larger than 9 eV, the  differences between the calculated and the experimental $d_{min}$ and $d_{max}$ are less than 3 \%.  Previous papers have chosen the $U$ value $>$ 6.5 eV for Cu atom in copper oxides case\cite{Anisimov91,Ekuma14,Nolan06,Elfimov08,Mishra16}. Furthermore, to ensure the reliability of our $U_{eff}$ value, we performed the linear response method and obtained the self-consistent $U_{eff}$ for Cu atoms of 11.5 eV, which agrees well with our finding in Fig. \ref{dist}. 

In addition, we performed full relaxation using the non-CDW structure as in Fig. \ref{str}(a) and (b). The lattice parameter $c$ is reproduced well with the inclusion of van der Waals interaction and PBEsol functional as in Table \ref{table_latt}. However, other lattice parameters, $a$ and $b$, and atomic positions do not considerably depend on the simulation conditions. Also, the scheme dependence of van der Waals correction is not significant. All calculated lattice parameters and atomic positions are comparable to the experimental values regardless of the condition. 


\begin{center}
\begin{table}[t]
\begin{ruledtabular}
\begin{tabular*}{0.5\textwidth}{@{\extracolsep{\fill}} c c c c c c c c}
& functional       & a             & b             & c            &z$_{Cu}$      &z$_{Te}$      &\\
\hline                                                                           
& PBE              & 3.280  & 4.018  &7.457 & 0.466 & 0.242 &\\  
& PBEsol           & 3.197  & 3.949  & 6.921 & 0.463 & 0.222 &\\  
& PBE+vdW          & 3.170  & 3.996  & 6.919 & 0.457 & 0.219 &\\  
& PBE+D3+U9    & 3.081  & 4.035  & 6.986 & 0.455 & 0.220 &\\  
& PBE+D2+U9    & 3.093  & 4.338  & 6.950 & 0.452 & 0.219 &\\  
& PBE+D3-BJ+U9   & 3.074  & 4.010  & 6.830 & 0.455 & 0.214 &\\  
& PBE+U9        & 3.138  & 4.101  & 7.415 & 0.459 & 0.236 &\\  
\hline                                                                                       
& EXP$^*$          & 3.138         & 4.059         & 6.902        & 0.454        & 0.221        &\\
\end{tabular*}
\end{ruledtabular}
\caption{Calculated lattice parameters and atomic positions of the non-CDW phase depending on the simulation condition. U9 means the $U_{eff}$ of 9eV.
}
\label{table_latt}
\end{table}
\end{center}

\begin{figure}[t]
\includegraphics[width=8.5cm]{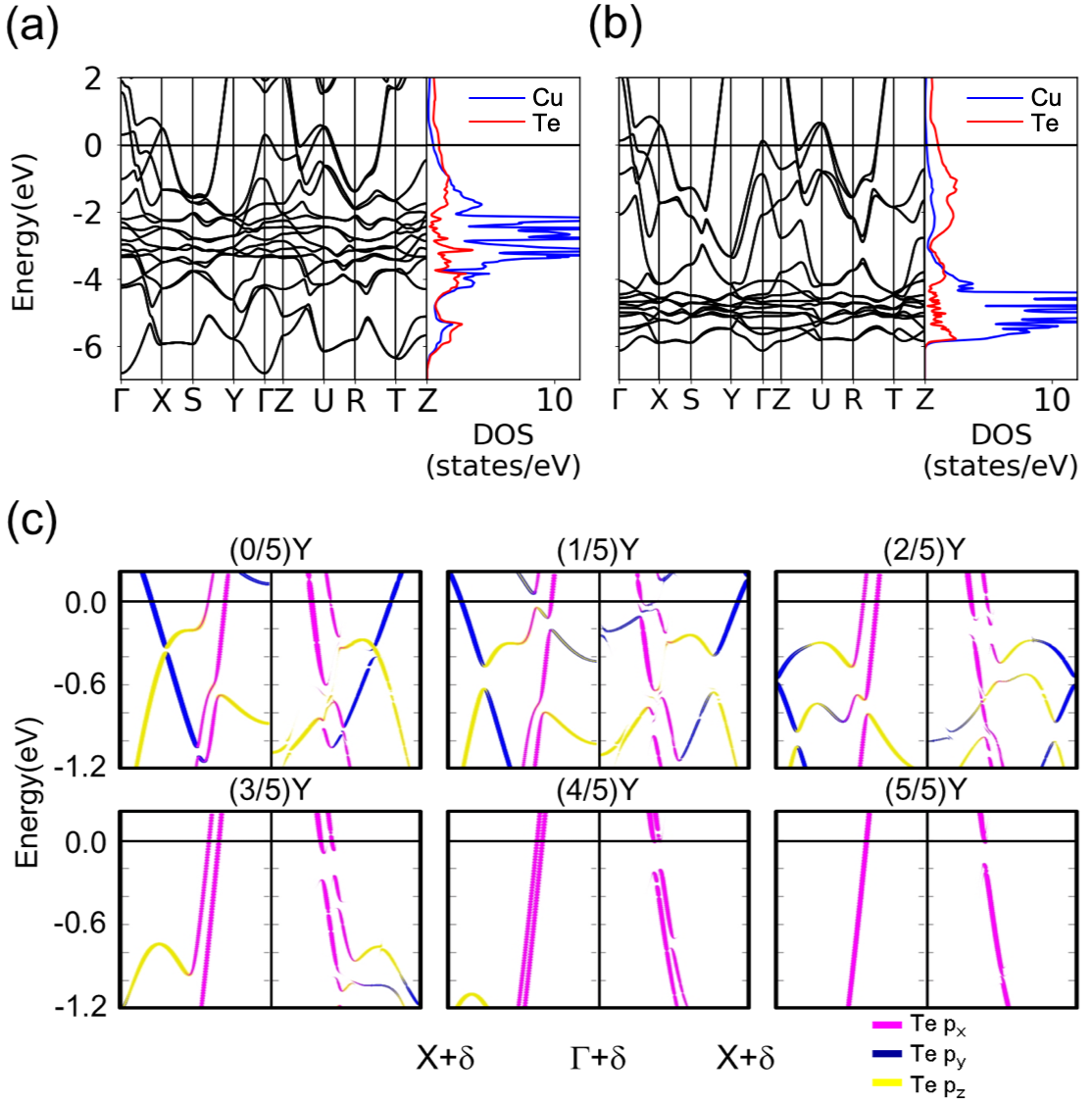}
\caption{Band structures and DOS of the non-CDW structure with (a) PBE and (b) PBE+$U$. (c) Band unfolding data with the CDW structure. On each figure, a mirror image of high symmetry BS (left) is compared with CDW BS (right).
}
\label{band}
\end{figure}

The correlation effect of Cu $d$ orbitals is investigated by observing the band dispersion and density of states (DOS) without(a) and with(b) $U$ as in Fig. \ref{band}. The band dispersion dominated by Te $p_x$ ($\sigma$ bond along the $a$-axis) and $p_y$ ($\pi$ bond along the $a$-axis) near the Fermi level, $E_f$ is hardly affected by the inclusion of $U$, however, the strong Cu weight redistribution to higher binding energy centering -5eV is observed, which suggests a non-negligible modification in Cu-Te hybridization near the $E_f$. As a result, Te character becomes more pronounced at the $E_f$ 
as in Fig. \ref{band}(b). The Cu weight shift from Fermi level can strengthen the 1D nature by removing Cu-Te hopping channel, which is a suitable condition for the CDW transition.
Note that Cu bands are located in the range of -2 to -4 eV, and in the range of -4 to -6 eV in PBE and PBE+$U$ calculations, respectively. Thus, the experimental measurement of Cu weight might be interesting to check the correlation effect of Cu.

Figure \ref{band}(c) shows the modification in band structures with the Te modulation: six figures present unfolded band structures along $\Gamma$-X shifted by $\delta = \alpha$Y  ($\alpha$= 0, 1/5, 2/5, 3/5, 4/5, and 1), as in the supplementary of Ref. \onlinecite{Zhang18}.
The most announced change occurs in the Te $p_x$ channel as expected. The band gap starts to open when $\alpha$ $>$ 2/5, which is consistent with the gap size dispersion along $k_y$ in the previous experiment\cite{Zhang18}. 
The overall feature of Te weight agrees well with the experimental observation that the CDW band gap formed by Te $p_x$ orbitals. In addition, our and previous unfolding data\cite{Zhang18} require a slight shift in energy to match with the experimental results. This need for the shift in energy might come from the hole doping on the system. However, the CDW structure with hole doping was restored to the high symmetric non-CDW structure after structural relaxation without the inclusion of $U$, which discards the doping-derived CDW transition scenario. 

\begin{figure}[t]
\includegraphics[width=8.5cm]{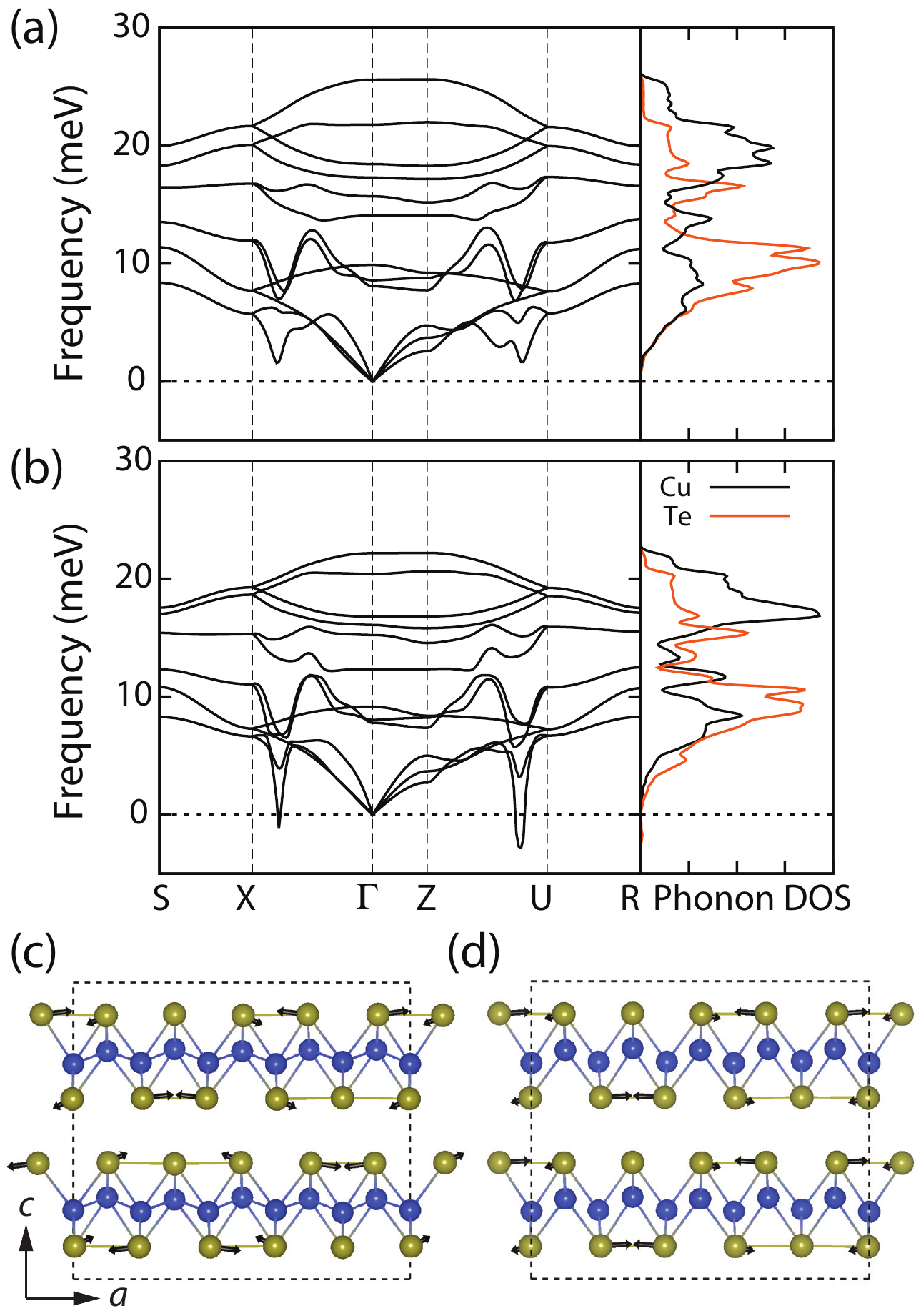}
\caption{Phonon dispersion curves and phonon DOSs of CuTe: (a) result from the PBE and (b) result from the PBE+$U$. The imaginary phonon frequencies implies the structural instability. Lattice displacements by the phonon soft modes at (c) $q_{ph0}=(0.4, 0.0, 0.5)$ and (d) $q_{ph1}=(0.4, 0.0, 0.0)$.
}
\label{ph}
\end{figure}
Figure \ref{ph}(a) and (b) show the phonon dispersion curves of the non-CDW state using PBE and PBE+$U$, respectively. We did not include the SOC for the PBE case to compare with previous phonon calculations\cite{Zhang18,Gamboa18}. The phonon structures of PBE+$U$ without SOC and PBE+$U$+D3 are qualitatively the same with that of PBE+$U$ in Fig. \ref{ph}(b), which show the imaginary phonon frequencies at  $q_{ph0}=(0.4, 0.0, 0.5)$ and at $q_{ph1}=(0.4, 0.0, 0.0)$. The imaginary phonon frequency so-called phonon soft mode indicates the structural instability. This demonstrates the critical role of the Coulomb correlation of Cu $d$ electrons in the CDW, which is consistent with structure relaxation results shown above. The phonon bands soften with the addition of the Coulomb correlation. Especially, as in the phonon DOSs of Fig. \ref{ph}, while the Te and Cu phonon bands in the PBE result are similarly occupied at a low frequency range below 5 meV, the Te bands are more occupied and Cu bands are less occupied at the low frequency range in the PBE+$U$ calculation. 
The phonon DOSs show that with inclusion of $U$, Cu weight becomes decoupled in the low frequency region where the soft mode is located, which supports our weakened Cu-Te bonding scenario from the electronic DOS analysis.
It is worth noting that the correlation of Cu $d$ adjusts not only Cu phonon bands but also Te phonon bands, leading to the imaginary phonon frequency of the Te phonon bands. The non-CDW structure exhibits a stable phonon dispersion curve in Fig. \ref{ph}(a) (also reported by Zhang {\it et al.}\cite{Zhang18}) without consideration of the Coulomb correlation despite the experimentally unstable non-CDW structure at a low temperature. The correlation-assisted phonon soft mode and structural transition as in Fig. \ref{ph}(b) has been reported in similar quasi-1D systems\cite{Kim13,Kim14}. 

The phonon instability at $q_{ph0}$ reproduces the the experimental $q_{CDW}$. Figure \ref{ph}(c) illustrates the lattice displacements of the softened phonon mode at $q_{ph0}$. This displacement of Te atoms generates 5$\times$1$\times$2 modulation of Te chains, which are not captured in the previous phonon calculations\cite{Zhang18,Gamboa18}. Our results demonstrate the electron-electron correlation and electron-phonon coupling play an essential role in driving the CDW. In addition, Figure \ref{ph}(d) represents the corresponding lattice displacement of the phonon soft mode at $q_{ph1}$. It also contains the same Te-Te modulation in a layer but does not change along the $c$ direction. The relaxed structure from $q_{ph0}$ modulation has lower energy of 1.6 meV than that from $q_{ph1}$ modulation. It explains why the CDW occurs at $q_{ph0}$ not at $q_{ph1}$.  The phonon instability is mostly related to the Te quasi-1D chain. And the small energy difference between $q_{ph0}$ and $q_{ph1}$ is related to chain-chain interaction which is the second order effect.

\section{Conclusions}
In conclusion, we demonstrated that using DFT and phonon analysis, the Coulomb correlation of Cu 3$d$ orbital plays an indirect but crucial role in the CDW transition of Te chains in the layered CuTe. 
We found that the inclusion of $U$ pushes away Cu $d$ orbitals from $E_f$ to the higher binding energy region and accordingly weakens the Cu-Te bonding. This strengthens 1D nature of Te-Te boning in the Te chain resulting in the CDW instability. 
Only with the inclusion of the Coulomb correlation in Cu atom, we observed the experimentally consistent imaginary phonon soft mode whose corresponding lattice distortion reproduces the CDW modulation. We believe that our work can shed light on the understanding of a mechanism of CDW transitions, especially when the interplay of multiple physical parameters are on effect.

\begin{acknowledgments}
We acknowledge the fruitful discussion with J. H. Shim and K.-T. Ko.
This work was supported by the NRF Grant 
(Contracts No. 2016R1D1A1B02008461, No. 2017M2A2A6A01071297, No. 2018R1D1A1A02086051),
Max-Plank POSTECH/KOREA Research Initiative (Grant No. 2016K1A4A4A01922028),
and the KISTI supercomputing center (Project No. KSC-2018-CRE-0075, KSC-2018-CRE-0079).
\end{acknowledgments}


\end{document}